\documentclass[final,5p,times,twocolumn]{elsarticle}

\journal{arXiv}

\usepackage{graphicx}
\usepackage{lineno,hyperref}
\usepackage[font=footnotesize, skip=8pt]{caption}
\usepackage{titlesec}
\titlespacing*{\section}{0pt}{*2.5}{*2}

\usepackage{filecontents}

\begin{document}

\begin{frontmatter}

\title{High resolution filtering and digitization system for cryogenic bolometric detectors}

\author[UNIMIB,INFN]{P. Carniti\corref{mycorrespondingauthor}}
\cortext[mycorrespondingauthor]{Corresponding author}
\ead{paolo.carniti@mib.infn.it}

\author[UNIMIB,INFN]{C. Gotti}

\author[UNIMIB,INFN]{G. Pessina}

\address[UNIMIB]{University of Milano-Bicocca, Piazza della Scienza 3, 20126, Milan (Italy)}
\address[INFN]{INFN of Milano-Bicocca, Piazza della Scienza 3, 20126, Milan (Italy)}

\begin{abstract}

BiDAQ is a custom filtering and data acquisition system designed for next-gen bolometric experiments dedicated to the search of neutrinoless double beta decay.
The system is composed of 12-channel analog-to-digital boards interfaced with FPGA SoC modules that collect and transmit the 24-bit data to the storage computers with a sampling rate up to 25 kHz.
Low noise, low power, high modularity and configurability, and an easy interface with Gigabit Ethernet are the main characteristics of the BiDAQ system.
%In this work we will present a general overview of the system, with emphasis on the FPGA back-end.

\end{abstract}

\begin{keyword}
Data acquisition system \sep FPGA \sep System-on-Chip \sep neutrinoless double beta decay \sep cryogenic calorimeter
\end{keyword}

\end{frontmatter}

%\linenumbers

\section{Introduction}
 
Next-generation cryogenic bolometric detectors, like those used by the CROSS \cite{CROSS} and CUPID \cite{CUPIDpreCDR} experiments for the search of neutrinoless double beta decay, will identify the type of interacting particles by measuring the amount of scintillation light produced in the crystals.
Each event will release most of its energy as heat in the crystal but also as light, captured by the light detectors facing the crystal.
Light detectors are made by Silicon or Germanium slabs that absorb photons and convert them to phonons, then measured with the same bolometric technique as the main heat signal, using thermistors.
Light detectors, however, are much lighter and are characterized by smaller and faster signals then those generated in the crystals.
They will thus require improved read out electronic systems with respect to previous generation experiments like CUORE \cite{Arnaboldi2018}, in particular with a higher bandwidth (from 100~Hz of CUORE, up to several hundred Hz), resolution (from 18 to 24 bits), and channel count (from 988 to 3306 channels), and lower power consumption (factor 5).
The faster response of light detectors is also beneficial in order to reject pile-up events like those originating from the two-neutrino double beta decay, which is a large source of background in large mass crystals based on $\mathrm{^{100}Mo}$.

\section{Filtering and digitization system}

\begin{figure}
	\centering
	\includegraphics[width=\linewidth]{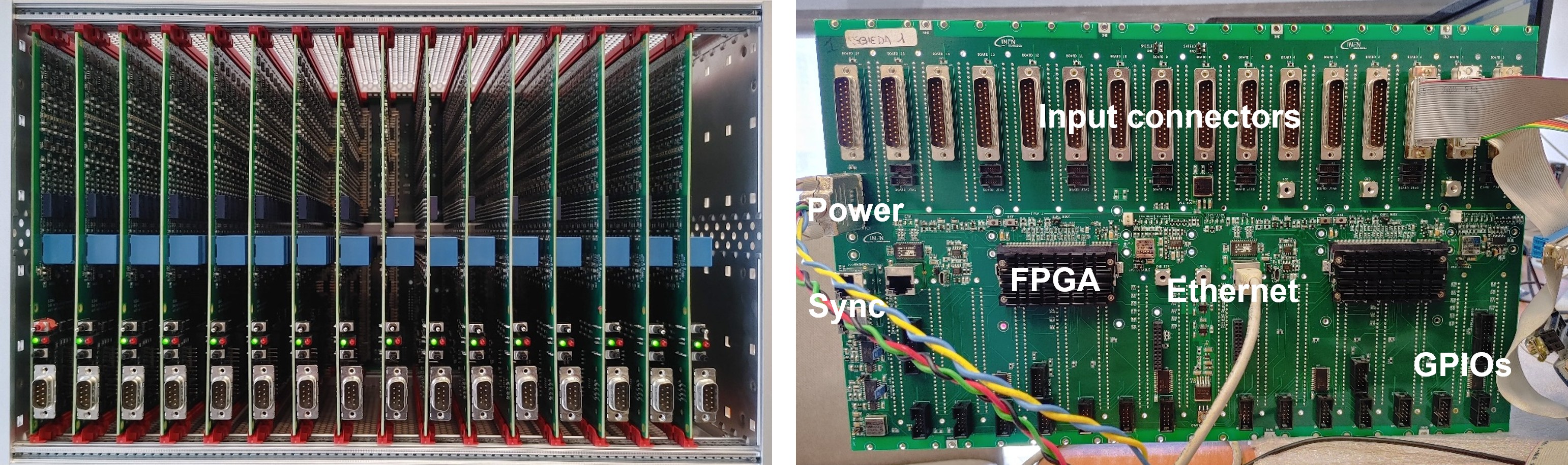}
	\caption{Front and rear view of the BiDAQ crates, fully populated with 16 boards (192 channels) and 2 FPGA modules.}
	\label{fig:crate}
\end{figure}

This paper will descrive the signal filtering and digitization system, named BiDAQ, for these experiments. 
BiDAQ is based on a custom solution comprised of several analog-to-digital boards interfaced to Altera Cyclone V SoC FPGA modules installed on the backplane of the BiDAQ crates, shown in Figure \ref{fig:crate}.
Each analog-to-digital board hosts 12 channels that allow signal digitization up to 25 ksps per channel and an effective resolution of 21 bits at 5 ksps, using 2-channels $\Delta$$\Sigma$ ADCs (Analog AD7175-2).
The anti-aliasing filter has a 6-pole Bessel-Thomson frequency response and the cut-off frequency can be digitally adjusted with 10 bits of resolution from 24 Hz to 2.5 kHz and thus adapted to the characteristics and working point of each detector, being it a light or heat one.
More details on the characteristics and performance of the boards are summarized in a previous work \cite{crossdaq}.

\section{FPGA modules}

The SoC FPGA modules installed on the backplane have a three-fold function: they control all the acquisition and board parameters through a Python-based server running on the FPGA SoC; they are responsible for the synchronization of the various analog-to-digital boards on the same rack and across different racks; they manage the transfer of the digitized data stream to the storage.

The used modules are commercial ones from Enclustra (Mars MA3), hosting inexpensive Altera Cyclone V FPGAs with a dual core ARM processor running Embedded Linux.
The modules are very compact (SO-DIMM form factor) but feature a sufficiently high number of IOs to collect data from all the ADCs, and Gigabit Ethernet connectivity towards the back-end.

Each module, out of the two installed on the crate backplane, collects data from 8 boards (96 channels), for a total of 16 boards per crate (192 channels).
The measured power consumption of the full system is 250 mW/channel with a 0 V DC input signal and 330 mW/ch with a 5 V one.
This represents at least a factor 5 improvement over CUORE DAQ system \cite{Domizio_2018}.

\subsection{Acquisition control}

The BiDAQ system is highly configurable.
Each channel can be individually enabled, its cut-off frequency can be adjusted, the inputs can be grounded, plus many other settings.
Several parameters of the FPGA can also be changed (sampling frequency, synchronization method, etc.), and embedded routines can be executed (DAQ start, DAQ stop, calibrations, monitoring, etc.).
All these slow controls can be provided using Python-based daemons running on the FPGA SoCs.
These daemons use the MQTT IOT messaging protocol \cite{mqtt} to receive commands from the central DAQ computer which runs the main DAQ software.
The daemons then executes methods from a Python class, which interacts with the single boards through CAN bus interface, and with the internal FPGA settings through memory-mapped registers.

\subsection{Synchronization}

The synchronization between the ADCs is accomplished by providing a common SYNC signal that drives the start of each ADC conversion.
The period of this signal determines the sampling frequency.
Since the SYNC is in common between all the ADCs on a single board, the 12 channels must share the same frequency.
Each ADC runs with the internal clock source and synchronization is granted by this common start.

An alternative synchronization method is available, although not yet implemented in firmware.
This will provide a common clock to all the ADCs and will use the SYNC to synchronize the start of the acquisition.
With this method it will be possible to set a different sampling frequency every 2 channels.

In both methods, data is downloaded from the ADC using 20 MHz SPI lines, one per ADC, for a total of 48 lines per FPGA.

The different FPGAs are then synchronized using distributed (daisy-chained) lines: a clock reference (max 16 MHz) and an open-drain start signal.
These lines can be either generated by a master FPGA, or come from an external device and are distributed across backplanes using differential lines over dedicated Ethernet cables.
Any FPGA can become master by simply issuing a specific command through the slow control.

\subsection{Data transfer}

\begin{figure}
	\centering
	\includegraphics[width=.75\linewidth]{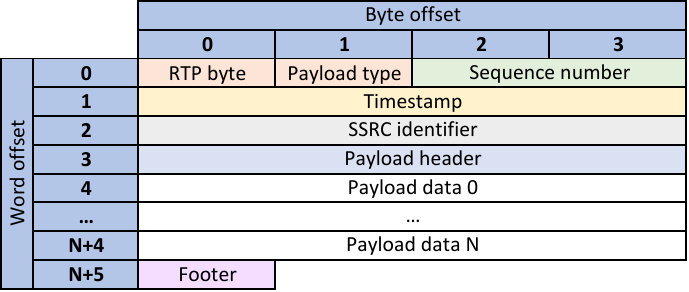}
	\caption{Packet format (Ethernet, IP and UDP headers are omitted).}
	\label{fig:packet}
\end{figure}

In cryogenic bolometric experiments, the data rate and the number of channels are not huge (about 3k channels for CUPID, corresponding to 60 MB/s at 5 ksps), hence it is common practice to store the continuous data stream and apply the trigger algorithms off-line.
After being collected by the FPGA, the data is transmitted to the storage over a standard 1 Gbps Ethernet.
The data is formatted by the FPGA firmware in UDP packets, using the RTP (real-time transport protocol) format \cite{rtp}.
This packet format, used for audio-video streaming applications (VoiP, videoconferencing, etc.), implements several characteristics that are very useful, as shown later.
A diagram of the packet format, including the RTP header, is shown in Figure \ref{fig:packet}.

Each RTP packet is made of a header, followed by the actual data payload.
The sequence number field (green in Figure \ref{fig:packet}) hosts a progressive packet number that is used to detect packet drops or to re-order the packet sequence.
The timestamp field (yellow) contains the time of the first sample included in the payload.
The SSRC field (grey) includes the unique ID of the channel that is generating the data.
The RTP stream, thus, is made of several streams each one coming from a specific channel, identified by the SSRC.
The payload then contains an additional 32-bit header (blue), which specifies the sampling frequency and the data format, and then the acquired samples in 32-bit format (white), 24 bits for the ADC conversion and 8 bits for other auxiliary flags.
An 8-bit footer (purple) at the end of the payload stores flags regarding hardware or firmware errors and can be monitored to detect data acquisition problems.

\subsection{Other functions}

Each FPGA module provides an 8-bit general purpose IO port that can be used to acquire additional data that must be synchronized in the acquisition stream.
Each bit of the port can be enabled and configured as output or input.
These signals can be used, for example, to trigger external devices, like the pulsers used for detector stabilization, or to read auxiliary digital flags, like those coming from a muon veto system.

The backplane can also host an add-on board with a microcontroller in order to supervise the FPGA modules, monitor the power supplies, and extend the slow control interface between the modules and the front-end boards, either through optically decoupled CAN bus or I2C.

\section{Conclusions and prospects}

The BiDAQ system is in an advanced phase of development and is currently operating in several CROSS and CUPID test facilities in Canfranc, LNGS, Milan and Orsay.

\bibliography{\jobname}

\begin{thebibliography}{1}
\expandafter\ifx\csname url\endcsname\relax
  \def\url#1{\texttt{#1}}\fi
\expandafter\ifx\csname urlprefix\endcsname\relax\def\urlprefix{URL }\fi
\expandafter\ifx\csname href\endcsname\relax
  \def\href#1#2{#2} \def\path#1{#1}\fi

\bibitem{CROSS}
I.~Bandac, et~al., {The 0$\nu$2$\beta$-decay CROSS experiment: preliminary
  results and prospects}, JHEP 2020 (01 2020).
\newblock \href {https://doi.org/10.1007/JHEP01(2020)018}
  {\path{doi:10.1007/JHEP01(2020)018}}.

\bibitem{CUPIDpreCDR}
{The CUPID Interest Group}, {CUPID pre-CDR} (2019).
\newblock \href {https://doi.org/10.48550/arxiv.1907.09376}
  {\path{doi:10.48550/arxiv.1907.09376}}.

\bibitem{Arnaboldi2018}
C.~Arnaboldi, et~al., A front-end electronic system for large arrays of
  bolometers, JINST 13~(02) (2018) P02026.
\newblock \href {https://doi.org/10.1088/1748-0221/13/02/p02026}
  {\path{doi:10.1088/1748-0221/13/02/p02026}}.

\bibitem{crossdaq}
P.~Carniti, et~al., {High-Resolution Digitization System for the CROSS
  Experiment}, {J Low Temp Phys} 199 (2020) 833--839.
\newblock \href {https://doi.org/10.1007/s10909-019-02249-9}
  {\path{doi:10.1007/s10909-019-02249-9}}.

\bibitem{Domizio_2018}
S.~D. Domizio, et~al., {A data acquisition and control system for large mass
  bolometer arrays}, {JINST} 13~(12) (2018) P12003.
\newblock \href {https://doi.org/10.1088/1748-0221/13/12/p12003}
  {\path{doi:10.1088/1748-0221/13/12/p12003}}.

\bibitem{mqtt}
\href{https://mqtt.org}{{MQTT, The Standard for IoT Messaging}} (2022).
\newline\urlprefix\url{https://mqtt.org}

\bibitem{rtp}
\href{https://datatracker.ietf.org/doc/html/rfc3550}{{RFC 3550, RTP: A
  Transport Protocol for Real-Time Applications}} (2003).
\newline\urlprefix\url{https://datatracker.ietf.org/doc/html/rfc3550}

\end{thebibliography}

\end{document}